# THE IMPORTANCE OF RANDOM GALILEAN TRANSFORMATION INVARIANCE IN MODELLING DISPERSED PARTICLE FLOWS

Michael W. Reeks
Nuclear Electric Company
Berkeley Nuclear Laboratories
Berkeley, Gloucestershire, United Kingdom

Sean McKee
Department of Mathematics
University of Strathclyde
Glasgow, United Kingdom

## ABSTRACT

The principle of Random Galilean Transformation (RGT) Invariance is applied to the random motion of particles in a turbulent gas to construct a kinetic equation for the transport of the particle phase space probability $< X, t) >$ where u and X are the velocity and position of a particle at time t. The essential problem is to find closed expressions for the phase space diffusion current $< Wf >$ where is the fluctuating aerodynamic force at u, x, t. The simplest form consistent with RGT Invariance, the correct equation of state and form of the interphase momentum transfer term is shown to be

$$< \underline{f} W > = - \left( \underline{\mu} \cdot \frac{\partial}{\partial \underline{v}} + \underline{\lambda} \cdot \frac{\partial}{\partial \underline{x}} \right) < W >,$$

in which $\underline{\mu} = < \underline{f}(t)\underline{v}(t) >$ and $\underline{\lambda} = < \underline{f}(t)\underline{x}(t) >$. This approach to modelling gas-solid flows is currently being used to investigate the behaviour of radioactive aerosols inside gas-cooled nuclear reactors.

## INTRODUCTION

This paper is concerned with the use of certain important invariance principles and properties associated with the motion of discrete particles in a random flow field, to construct a kinetic equation for a dilute dispersed phase of particles in a turbulent gas. The kinetic equation to which we refer, is the averaged Liouville equation for the dispersed phase and represents the transport of the particle phase space probability. In this particular problem, the phase space probability is denoted by $< W(u, X, t) >$ and refers to the probability density that a particle has a velocity $\underline{v}$ and position at time t; W means the phase space density for one realisation of the random flow field, and $< ... >$ an ensemble average over all realisations of the flow field. The equation for $< W(u, X, t) >$ is for example the analogue of the Boltzmann equation of Classical Kinetic Theory and can be used in a similar way to construct the so-called continuum equations and constitutive relations of a two-fluid model of a dispersed particle flow whilst also providing the essential framework in which the boundary conditions appropriate for the solution of those equations can be resolved. This approach is currently being investigated as a method for calculating the retention of radioactive gas-borne particles accidentally released into the coolant circuit of a gas-cooled reactor. This retention depends critically upon the adhesion of a particle on impact with a surface and hence upon the distribution of particle impact velocities.

If the underlying equation for the particle random motion is a Langevin equation of the form:

$$\frac{d\underline{u}}{dt} - \beta \underline{v} = \underline{f}(t),  \quad (1)$$

then the kinetic equation for the transport of W has the form:

$$\left( \frac{\partial}{\partial t} + \underline{v} \cdot \frac{\partial}{\partial \underline{x}} - \beta \frac{\partial}{\partial \underline{v}} \cdot \underline{v} \right) < W(\underline{v}, \underline{x}, t) >$$
$$= - \frac{\partial}{\partial \underline{v}} \cdot < \underline{f}(\underline{x}, t) W(\underline{v}, \underline{x}, t) > \quad (2)$$

where in equations (1) and (2), is the inverse particle response time, and $(X, t)$ is the fluctuating aerodynamic driving force (per unit mass). f(t) in equation (1) specifically refers to the aerodynamic force along a particle trajectory i.e.

$$\underline{f}(t) = \underline{f}(\underline{x}(t), t),  \quad (3)$$

where x(t) is the solution of equation (1) for some arbitrary initial conditions. Equation (2) is an unclosed equation in that $<(W >$ is unknown. The construction of a kinetic equation is therefore related to the problem of finding suitable forms for the phase space diffusion current $<$ in terms of $< W >$ and the moments of f(x, t) If f(t) is assumed to be white noise i.e. $\tau$ the integral timescale of ( is $<< \beta$ the problem is relatively straightforward. In this case the motion of an individual particle is a Markov Process, for which the Chapman-Kolmogoroff equation is appropriate i.e.
the phase space diffusion current reduces to the form used in the classical Fokker Planck equation (CFP), namely

$$< W \underline{f} > = - \underline{\mu} \cdot \frac{\partial}{\partial \underline{v}} < W >, \quad (4)$$

where e is given by

$$\underline{\mu} = \int_0^\infty < \underline{f}(0)\underline{f}(s) > ds \quad (5)$$

for $t >> \tau$. This is the form used by Buyevich (1972) in his treatment of dispersed particle flows. However we wish to concern ourselves here with the more general case when the white noise limit does not necessarily apply.

> To deal with this general case we have appealed here to an important invariance property not satisfied by the classical Fokker Planck equation (and by implication the Chapman-Kolmogoroff equation) when the restriction on timescale is lifted, namely invariance to a random Galilean transformation (RGT). The terminology and usage is due to Kraichnan (1977) and means applying to each realisation of the carrier flow a translational velocity, constant in space and time but varying randomly in value from one realisation to the next; in Kraichnan's usage of RGT the distribution of velocities is taken to be Gaussian for convenience. Clearly the internal dynamics should be unaffected by this transformation and should be reflected in the equations that describe the average behaviour of the resulting system. In the case of the CFP equation the terms that describe the dispersion due to the aerodynamic driving force and that due to the translational velocity should be separate. Whilst this separation does occur the form of the dispersion term due to the random translational velocity is incorrect. More precisely this failure to preserve RGT invariance means failure to reproduce the correct equation of state for the dispersed phase.

The principal objective here will be to derive the simplest possible kinetic equation which preserves RGT invariance and in turn:

- is consistent with the correct equation of state of the dispersed phase for all particle response times in homogeneous stationary flows;
- contracts to the CFP equation in the white noise limit.

We shall see that these conditions are met at the simplest level by a phase space diffusion current proportional to both velocity and spatial gradients in the particle phase space probability. The method of solution logically follows from an application of RGT, and leads to a formal expansion of the phase space diffusion current in successively higher order cumulants of the aerodynamic driving force along a particle trajectory. The simplest form of this diffusion current corresponds to the first term in the expansion. More important it corresponds to the case when the aerodynamic force is a Gaussian random process, for which all the terms in the expansion except the first term contract to zero.

RGT INVARIANCE

As we stated in the introduction the particular form for the phase space diffusion current given in equation (5) does not preserve RGT invariance. To show this let us apply to each realisation of the system a constant velocity $u_0$, random from one realisation to the next and statistically independent of For convenience we shall assume the distribution of is isotropic and Gaussian. To avoid unnecessary complications we shall suppose that Stokes drag holds so that adding a uniform translational velocity to each realisation of the carrier flow is equivalent to adding a term $\beta u_0$ to the right-hand side of the equation of motion (1). Transforming to the variables

$$\underline{v} - \underline{v}_0 (1 - e^{-\beta t}), \quad \underline{x} - \underline{v}_0 \beta^{-1}(1 - e^{-\beta s}) \quad (6)$$

leaves the original equation of motion in these new variables unchanged. Thus if $<W>^{(1)}$ is the new average phase space density

$$<W>^{(1)} = <\,<W(\underline{v} - \underline{v}_0(1 - e^{-\beta t}),\, \underline{x} - \underline{v}_0\beta^{-1}(1 - e^{-\beta s})ds,\, t)>^{(0)}\,> \quad (7)$$

where $<W(\underline{v}, \underline{x}, t)>^{(0)}$ is the average phase space density in the absence of the applied random velocity. Expanding the right-hand side in terms a Taylor-series gives formally

$$<W>^{(1)} = <\exp(-\underline{v}_0 \cdot ((1 - e^{-\beta t})\frac{\partial}{\partial \underline{v}} + \int_0^t (1 - e^{-\beta s})ds\frac{\partial}{\partial \underline{x}}))> \cdot <W(\underline{v}, \underline{x}, t)>^{(0)} \quad (8)$$

The characteristic function $M(\underline{k})$ of is given by

$$M(\underline{k}) = <\exp(i\underline{k} \cdot \underline{v}_0)> \quad (9)$$

So $<W>^{(1)}$ can be written formally as

$$<W>^{(1)} = M(\underline{k}) \{(1 - e^{-\beta t})\frac{\partial}{\partial \underline{v}} + \int_0^t (1 - e^{-\beta s})ds\frac{\partial}{\partial \underline{x}}\} <W>^{(0)} \quad (10)$$

Furthermore

$$<\underline{v}_0 W> = -i\frac{\partial}{\partial \underline{k}} M(\underline{k}) \Big|_{\underline{k} = i(1 - e^{-\beta t})\frac{\partial}{\partial \underline{v}} + i\int_0^t (1 - e^{-\beta s})ds\frac{\partial}{\partial \underline{x}}} <W>^{(0)} \quad (11)$$

For an isotropic Gaussian distribution

$$M(\underline{k}) = \exp(-\frac{1}{6}<v_0^2>k^2) \quad (12)$$

So from equations (11) and (10) we obtain

$$\beta <\underline{v}_0 W> = -\frac{1}{3}\beta<v_0^2>[(1 - e^{-\beta t})\frac{\partial}{\partial \underline{v}} + \int_0^t (1 - e^{-\beta s})ds\frac{\partial}{\partial \underline{x}}] <W>^{(1)} \quad (13)$$

Thus RGT invariance means adding the terms on the right-hand side of equation (13) to the right-hand side of equation (4). Formal application of the Chapman-Kolmogoroff equation in these circumstances is equivalent to replacing by $f + \beta u_0$ in the CFP form for in equation(5). Such a substitution does not generate correctly either of the velocity or spatial gradient terms in equation (13). In fact the spatial gradient term is not generated at all whilst the velocity gradient term is infinite in value. Whilst this failure to preserve RGT invariance is not surprising in view of the fact that RGT is not a Markov process, it does serve to demonstrate the total inadequacy of the CFP equation in cases where the random driving force is very far from the white noise limit. We see the second term in brackets on the right-hand side of equation (13) is vital in that it dominates the phase space diffusion current when $t \to \infty$. Clearly the random translational velocity will give rise in this instance to an extra spatial diffusion coefficient $(1/3)<v_0^2>t$ precisely accounted for by the inclusion of the second term in equation (13) in the phase space diffusion current.

let us now use the same approach to consider a simple random walk problem in which the driving force is perfectly correlated for time steps T whilst uncorrelated for times $>$ T. The problem is similar to the problem treated by Durbin (1980) for a simple model of fluid element dispersion. The motion approximates a Markov process for $\beta T \to 0$ whilst the case for $\beta T \to \infty$ corresponds to the case considered above. However we consider here the form of the phase space diffusion current for any value of $\beta T$. The solution of the equation of motion measured at time t from time zero, is

$$\underline{v}(t) - \underline{v}_0 e^{-\beta t} = \int_0^t \dot{g}(t - s) f(s)\, ds \quad (14)$$

$$\underline{x} - \underline{x}_0 - \underline{v}_0\beta^{-1}(1 - e^{-\beta t}) = \int_0^t g(t - s) f(s)\, ds$$

where and $x_0$ are the initial velocity and position at time zero and

$$g(t) = \beta^{-1}(1 - e^{-\beta t}). \quad (15)$$

Dividing the time interval t into a sequence of discrete times $t_j = j\tau$ for $j = 1, N$ where $N = t/\tau$, we define a corresponding sequence of random vectors $\underline{f}^{(j)}$ such that

$$\underline{f}(s) = \underline{f}^{(j)} \quad \text{for } t_{j-1} < s < t_j. \tag{16}$$

Thus

$$\underline{v}(t) - \underline{v}_0 e^{-\beta t} = \sum_{j=1}^{N} \phi_j(t) \underline{f}^{(j)}$$

$$\underline{x} - \underline{x}_0 - \underline{v}_0 \beta^{-1}(1 - e^{-\beta t}) = \sum_{j=1}^{N} \psi_j(t) \underline{f}^{(j)} \tag{17}$$

where

$$\phi_j(t) = \int_{(j-1)\tau}^{j\tau} \dot{g}(t-s) \, ds \ ; \ \psi_j(t) = \int_{(j-1)\tau}^{j\tau} g(t-s) \, ds. \tag{18}$$

Suppose we let $<W(\underline{v}, \underline{x}, t)>$ be the average phase space densities with and without the aerodynamic force $\underline{f}(t)$. Thus using similar arguments to those used above in considering RGT invariance, we obtain

$$<W> = <<W(\underline{v} - \int_0^t \underline{f}(s) e^{\beta(s-t)} ds, \underline{x} - \int_0^t \underline{f}(s)(1 - e^{-\beta(s-t)}) ds, t)^{(0)}>$$

$$= <\exp -[\sum_{j=1}^{N} \phi_j \underline{f}^{(j)} \cdot \frac{\partial}{\partial \underline{v}} + \psi_j \underline{f}^{(j)} \cdot \frac{\partial}{\partial \underline{x}}]> W^{(0)}(\underline{v}, \underline{x}, t), \tag{19}$$

where

$$W^{(0)}(\underline{v}, \underline{x}, t) = \delta(\underline{v} - \underline{v}_0 \dot{g}(t)) \delta(\underline{x} - \underline{x}_0 - \underline{v}_0 g(t)). \tag{20}$$

The characteristic function associated with the set of random variables $\underline{f}^{(j)}$ is given by

$$M[\underline{k}^{(1)}, \underline{k}^{(2)}, \ldots, \underline{k}^{(N)}] = <\exp[i \underline{k}^{(j)} \cdot \underline{f}^{(j)}]>. \tag{21}$$

So the right-hand side of equation (20) can be written as

$$<W> = M[i\phi_1 \frac{\partial}{\partial \underline{v}} + i\psi_1 \frac{\partial}{\partial \underline{x}}, \ldots, i\phi_N \frac{\partial}{\partial \underline{v}} + i\psi_N \frac{\partial}{\partial \underline{x}}] W^{(0)}, \tag{22}$$

and

$$<\underline{f}^{(j)} W> = -i \frac{\partial}{\partial \underline{k}^{(N)}} M[\underline{k}^{(1)}, \ldots, \underline{k}^{(N)}] W^{(0)} \tag{23}$$

with $\underline{k}^{(j)} = i\phi_j \frac{\partial}{\partial \underline{v}} + i\psi_j \frac{\partial}{\partial \underline{x}}$ for $j = 1, N$

If we assume the components of the random vectors $\underline{f}^{(j)}$ are such that

$$<f_i^{(l)} f_j^{(m)}> = <f_i f_j> \delta_{lm} \tag{24}$$

with a Gaussian characteristic function given by

$$<\underline{f} W> = -<\underline{f} \underline{f}> \cdot [\beta^{-1}(1-e^{-\beta\tau}) \frac{\partial}{\partial \underline{v}} + \beta^{-1}(\tau - \beta^{-1}(1-e^{-\beta\tau})) \frac{\partial}{\partial \underline{x}}] <W>. \tag{27}$$

For $\tau = t$ we generate the closure term for an applied random aerodynamic force (constant in time, the same as that given in equation (13) with $\tau = \Delta t$. At the other extreme with $\beta\tau < 1$ we note that the first term in brackets on the right-hand side will give rise to a global gaussian distribution of velocities at equilibrium with a variance $\sigma_v^2$ approaching $\tau/\beta$. Furthermore similar changes in $<W>$ for changes in $\sigma$ will occur for changes in $\sigma\beta^{-1}$. i.e. $\partial<W>/\partial x \sim \beta\partial<W>/\partial\sigma$. Thus comparing the first term in brackets on the right-hand side of equation (27) with the second we obtain

$$<\underline{f} W> = -\tau <\underline{f} \underline{f}> \cdot \frac{\partial}{\partial \underline{v}} (1 + O(\beta\tau)). \tag{28}$$

That is the form of the kinetic equation for $\beta\tau < 1$ contracts to the CFP equation.

This simple example and the application of RGT serve to illustrate the fact that at the simplest level the phase space diffusion current ought to be of the form

$$<\underline{f} W> = -\left(\mu \frac{\partial}{\partial \underline{v}} + \lambda \frac{\partial}{\partial \underline{x}}\right) <W>, \tag{29}$$

in order to preserve RGT invariance. In the case of the white noise approximation the spatial gradient contribution is lost because it is of order $\beta\tau$ smaller than the velocity gradient contribution. The white noise approximation is essentially the first term in a perturbation expansion for $<\underline{f}(\underline{x},t) W(\underline{v},\underline{x},t)>$. Only the second term in the expansion reveals any contribution from the spatial gradient. In this form the contribution has been noted by Van Kampen (1981) though its importance in non white noise problems was not appreciated. Only in the application of Kraichnan's Lagrangian History Direct Interaction (LHDI) Approximation (1977) to similar problems to the one considered here has its importance been recognised as a necessary consequence of preserving RGT invariance. We refer here explicitly to Orzag's (1968) application of LHDI to the stochastic acceleration of charged particles by a prescribed electric field (5). It is the application of LHDI to this problem that forms the basis of a subsequent paper.

A CUMULANT EXPANSION FOR $<\underline{f} W>$

Let us now generalise the method used to obtain a closed expression for $<W>$ by considering $\underline{f}(t)$ as a random field in which $\underline{f}(t)$ is regarded as a continuous process represented by the limit of the discrete process

$$N \to \infty \, L[\underline{f}(s_1), \underline{f}(s_2), \ldots, \underline{f}(s_N)] \text{ where } s_j = j\tau,$$
$$\text{with } \tau = t/N. \tag{30}$$

More precisely $\underline{f}(s)$ refers to $\underline{f}(\underline{u}, \underline{l}, t)$ i.e. the aerodynamic force measured at time $s$ for $s \leq t$ along a particle trajectory which passes through the point $(\underline{u}, \underline{x})$ at time $t$. So

$$M[\underline{k}^{(1)},...,\underline{k}^{(N)}] = \exp[-\frac{1}{2} <f_i f_j> \sum_{l=1}^{N} k_i^{(l)} k_j^{(l)}] \quad (25)$$

Then

$$<\underline{f}W> = -<\underline{f}\,\underline{f}>\cdot[\phi_N \frac{\partial}{\partial \underline{v}} + \psi_N \frac{\partial}{\partial \underline{x}}]<W>. \quad (26)$$

Explicitly evaluating and we obtain finally

$$M[\underline{\phi}(s)] = \int ....\int d\underline{f}(s_1)...d\underline{f}(s_N) \exp(i\underline{\phi}^c \cdot \underline{f}(s_j)) P_N(\underline{f}(s_1)...\underline{f}(s_N))$$
with $\underline{\phi}^{(l)} = \underline{\phi}(s_j)\tau,$ (33)
and $N \to \infty$

where $P_{NC}$ (st), is the probability density for the occurrence of $[\underline{f}(s_1)....\underline{f}(s_N)]$. As before we have

$$<W> = <\exp\{-\int_0^t ds \underline{f}(s) \cdot [\dot{g}(t-s)\frac{\partial}{\partial \underline{v}} + g(t-s)\frac{\partial}{\partial \underline{x}}]\}>$$

which we can write as

$$<W> = M[\underline{\phi}(s)]W^{(0)} \quad (35) \text{ with } \underline{\phi}(s) = i[\dot{g}(t-s)\frac{\partial}{\partial \underline{v}} + g(t-s)\frac{\partial}{\partial \underline{x}}]$$

$$<\underline{f}(\underline{x},t)W(\underline{v},\underline{x},t)> = -i\frac{\delta M[\underline{\phi}(s)]}{\delta \underline{\phi}(t)}W^{(0)} \text{ with } \underline{\psi}(s) = \hat{\underline{\psi}}(t-s) \quad (36)$$

where $\delta/\delta \underline{\phi}(t)$ means a functional derivative for changes of $\underline{\phi}(s)$ about $s' = t$. Following e.g. Monin & Yaglom (1971) we now represent $M[\underline{\phi}(s)]$ in the form

$$M[\underline{\phi}(s)] = \exp[\sum_{n=2}^{\infty} \frac{i^n}{n!}\int_0^t ds_1....\int_0^t ds_n ||<f_{i_1}(s_1)...f_{i_n}(s_n)>|| \phi_{i_1}(s_1)....\phi_{i_n}(s_n)] \quad (37)$$

where $||<f_{i_1}(s_1)...f_{i_n}(s_n)>||$ is the cumulant associated with the average $<f_{i_1}(s_1)...f_{i_n}(s_n)>$. Substituting this in equation (36) we obtain

$$<\underline{f}W> = +\sum_{n=1}^{\infty} \frac{(-1)^n}{n!}\int_0^t ds_1....\int_0^t ds_n ||<f_{i_1}(s_1)...f_{i_n}(s_n)\underline{f}(t)>|| \hat{\psi}_{i_1}(t-s_1)....\hat{\psi}_{i_n}(t-s_n)<W(\underline{v},\underline{x},t)>. \quad (38)$$

$$\underline{f}(t) = \underline{f}(\underline{v},\underline{x},t|t) = \underline{f}(\underline{x},t). \quad (31)$$

The characteristic functional of $f(s)$ is given by

$$M[\underline{\phi}(s)] = <\exp i\int \underline{\phi}(s) \underline{f}(s)ds>, \quad (32)$$

where < > implies an average over all realisations of the random field $f(s)$, represented formally by an integration over the infinite set of discrete vectors i.e.
and
As equilibrium is approached in homogenous stationary flows, concentration gradients approach zero and thus

$$<\rho \underline{f}(\underline{x},t)> \to -<\underline{f}(t)\underline{x}(t)>\cdot\frac{\partial}{\partial \underline{x}}<\rho(\underline{x},t)>, \quad (42)$$

Multiplying the kinetic equation for $<W>$ by and integrating over all velocities with $t$ gives for the equation of state in homogeneous stationary flows

$$\frac{p}{<\rho>} = \frac{1}{3}\beta^{-1}\int_0^{\infty}<\underline{f}(0)\underline{f}(s)>ds, \quad (43)$$

where $p$ is the pressure tensor. This is consistent with the form derived by Reeks (1984), from which using the same arguments leads to, as $t \to \infty$

$$\frac{\partial <n>}{\partial t} = \frac{\partial}{\partial \underline{x}}\cdot \underline{\varepsilon}(\infty)\frac{\partial <n>}{\partial \underline{x}}. \quad (44)$$

PROPERTIES OF THE CUMULANT EXPANSION

It is clear from replacing $\underline{f}(t)$ by $\underline{f}(t) + $ in the expansion that RGT invariance is satisfied at each order of the expansion for any arbitrary distribution of the translational velocity $\underline{v}_0$. Furthermore if $f(t)$ is a Gaussian random process, every term in the expansion apart from the first is zero and we have identically

$$<\underline{f}(\underline{x},t)W(\underline{v},\underline{x},t)> = -[\underline{\mu}(t)\cdot\frac{\partial}{\partial \underline{v}} + \underline{\lambda}(t)\cdot\frac{\partial}{\partial \underline{x}}]<W(\underline{v},\underline{x},t)>, \quad (39)$$

where

$$\underline{\mu}(t) = <\underline{f}(t)\underline{v}(t)> \quad ; \quad \underline{\lambda}(t) = <\underline{f}(t)\underline{x}(t)>. \quad (40)$$

Further investigation by integrating the equation for $<W>$ over all $\underline{x}$, generates an equation for the global velocity distribution from which the equations for the velocity moments can be found by multiplying the equation by appropriate orders of the velocity components. Not surprisingly these equations are identical to those obtained from the equation of motion.

Turning to the interphase momentum transfer term we have

$$< \rho \underline{f}(\underline{x},t) > = \int < \underline{f}(\underline{x},t) W(\underline{v},\underline{x},t) > d\underline{v}$$

$$= \sum_{n=1}^{\infty} \frac{(-1)^n}{n!} || < \underline{f}(t)(\underline{x}(t) \cdot \frac{\partial}{\partial \underline{x}})^n > || < \rho(\underline{x},t) >.$$

(41)

$$E_{(CO)} = \beta^{-1} \int_0^\infty < \underline{f}(0)\underline{f}(s) > ds,$$

(45)

To show that the cumulant expansion for $W(u, x,t) >$ contracts to the CFP form for ßT 00 and we use the property of the cumulant

$|| < \underline{f}_{i_1}(s_1)....\underline{f}_{i_n}(s_n)\underline{f}(t) > ||$ that it is only effectively non-zero for $s_1, s_2, ..., s_n$ within a range $\tau$ of $t$. Thus the nth term in the expansion $\sim || < \underline{f}^{n+1} > || \tau^n \partial^n < W > /\partial v^n$. With $\partial^n < W > /\partial v^n \sim \sigma^{-n+1} \partial < W >/\partial v$ where $\sigma^2 \sim < \underline{f}^2 > \tau/\beta$, then the nth term in the expansion $\sim < \underline{f}^2 > \tau(\beta\tau)^{(n-1)/2}(\partial < W > /\partial v)$ i.e. for $\beta\tau \to 0$ and $t \gg \tau$

$$< \underline{f}(\underline{x},t) W(\underline{v},\underline{x},t) > = - \int_0^\infty \frac{< \underline{f}(0)\underline{f}(s) > ds}{[1 + O(\sqrt{\beta\tau})]} \cdot \frac{\partial < W >}{\partial \underline{v}}$$

(46)

The value of the cumulant expansion in equation (38) would be the hope that truncating the expansion at higher and higher order (cumulant neglect hypothesis) would achieve succesively better approximations for $W >$ for a non-gaussian f (t). Unfortunately the expansion is not uniformally

SUMMARY AND CONCLUDING REMARKS

The principal objective of this study has been the pursuit of an appropriate kinetic equation for a suspension of particles in a turbulent flow. We pointed out the value of this equation in deriving constitutive relations for the so called continuum equations for a two-fluid model of a dispersed particle flow.

Correct forms for the equation of state and fluctuating interphase momentum transfer term are basic features that a kinetic equation for the particle phase density must reproduce. In this respect the classical Fokker-Planck equation used by Buyevich (1972) prove inadequate outside the white-noise limit. This inadequacy we traced to a failure to preserve RGT invariance. In fact application of RGT gave rise to a method of constructing a formal expansion for the phase space diffusion current in higher order cumulants of the fluctuating aerodynamic driving force which satisfied RGT invariance at each order, and as a consequence gave correct forms for the equation of state and interphase momentum transfer term. For a Langevin equation of motion in which the driving force is a Gaussian process all terms in the expansion apart from the first contract to zero, giving for the phase space diffusion current

$$< fVV > = - \left( \underline{\mu} \cdot \frac{\partial}{\partial \underline{v}} + \underline{\lambda} \cdot \frac{\partial}{\partial \underline{x}} \right) < W >,$$

in which

$$\underline{\mu} = < \underline{f}(t)\underline{v}(t) > \quad \underline{\lambda} = < \underline{f}(t)\underline{x}(t) >.$$

This is the simplest form for the phase space diffusion current which reproduces the correct equation of state.

convergent: furthermore it is well known (4)that realisable solutions for

$< W >$ can only be achieved by truncating the expansion after the first term or including all the terms the in the expansion. Ultimately this non realisability can be traced to the use of a truncated cumulant expansion for the characteristic functional for f (t) which does not satisfy realisabilty for the coresponding probability functional for f(l) (6). However this does not rule out the use of such expansions. Whilst solutions for $< W >$ may be negative for some regions of phase space such regions may be regions where $< W >$ is in reality very small in value; over regions where $< W >$ is appreciable, better approximations can be obtained by taking more than one term in the expansion (Risken, 1986). However since the expansion is not uniformally convergent one has no a priori rule for assessing how many terms are required.

Ultimately we are limited by our knowledge off(t) as a random process. No exact theory exists for determining this process from the underlying random velocity field of the carrier flow from which it is derived. What does exist are approximate methods for determining the second moments off(t). Experimental determinations are similarly limited to measurements of this quantity. In view of these limitations it would be pointless from a practical point of view to consider currently more than the first term in the expansion which at least corresponds to a realisable process and one for which as we have shown becomes the dominant tenn in the expansion when f(t) behaves as white noise.

8. Il. Risken, "I'he Fokker-Planck Equation: Methods of Solution and Applications, Springer Verlag, 2nd. Ed., 1986

43

ACKNOWLEDGMENT

The work described was carried at Berkeley Nuclear Laboratories and is published with the permission of Nuclear Electric Co..